\documentclass[1,12pt]{elsarticle}
\usepackage{graphicx}
\usepackage{graphicx,epstopdf}
\usepackage{caption}
\usepackage{subcaption}
\usepackage{float}
\usepackage{amssymb,amsmath}
\usepackage{exscale}
\usepackage{makeidx,shortvrb,latexsym}
\usepackage{epstopdf}
\usepackage{tabularx, booktabs, multirow}
\usepackage{array}
\usepackage[]{hyperref}
\usepackage{color}

\newcounter{defcounter}
\begin{document}
\begin{frontmatter}

\title{Crystal plasticity simulation of the effect of grain size on the fatigue behavior of polycrystalline  Inconel 718}

\author{A. Cruzado$^{1, 2}$\corref{cor1}}
\author{S. Lucarini$^1, 3$}
\author{J. LLorca$^{1, 3}$\corref{cor1}}
\author{J. Segurado$^{1, 3}$}
\address{
$^1$ IMDEA Materials Institute \\ C/ Eric Kandel 2, 28906, Getafe, Madrid, Spain. \\\ \\
$^2$ Department of Aerospace Engineering, Texas A$\&$M University\\  H. R. Bright Building, 701 Ross St, College Station, TX 77840, USA.\\\ \\
$^3$ Department of Materials Science, Polytechnic University of Madrid/Universidad Polit\'ecnica de Madrid \\ E. T. S. de Ingenieros de Caminos. 28040 - Madrid, Spain. 
}
\cortext[cor1]{Corresponding authors at IMDEA Materials Institute, Spain.\\
 Email: acruzadogarcia@gmail.com (A. Cruzado), javier.llorca@imdea.org (J. LLorca)
}

\begin{abstract}

A microstructure-based model that accounts for the effect of grain size has been developed to study the effect of grain size on the fatigue life of Inconel 718 alloys. The fatigue behavior  of two alloys with different grain size was determined by means of uniaxial cyclic deformation tests under fully-reversed deformation ($R_\varepsilon$ = -1) at 400$^\circ$C in the low cycle fatigue regime.  The model was based in the determination of the fatigue indicator parameter (based on the local crystallographic strain energy dissipated per cycle) by means of computational homogenization of a representative volume element of the microstructure. The mechanical response of the single crystal within the polycrystal was modelled through a phenomenological crystal plasticity model which was modified to account for the effect of grain size on the monotonic and cyclic hardening/softening mechanisms. The microstructure-based crack initiation model parameters were calibrated from the experimental tests of the material with fine grain size.  The results of the fatigue simulations were in good agreement with the experimental results in terms of the cyclic stress-strain curves and of the number of cycles for fatigue crack initiation. The model did not show any grain size effect on the fatigue life for the largest cyclic strain ranges while  the predicted fatigue life predicted was considerably longer in the case of the microstructure with fine grain size for the lowest strain ranges, in quantitative agreement with experimental data. These differences were attributed to changes in the deformation modes between homogeneous plastic deformation at large cyclic strain ranges and localized plasticity in a few grains at low cyclic strain ranges.
\end{abstract}


\end{frontmatter}

\section{Introduction}

Inconel 718 is the most used superalloy in  turbine disks and other components working below 650$^{\circ}$C  because its relative low cost, exceptional low cycle fatigue (LCF) behavior, high strength, corrosion and creep resistance \citep{Locq2011}. The standard operation conditions are in the temperature range $400^{\circ}$C to $650^{\circ}$C \citep{BSP16} and design considerations impose the presence of  stress concentrations   in which the service life is controlled by the alloy response to LCF \citep{Shanyavskiy2013}. It is well known, from the qualitative viewpoint, that the strength and the LCF performance of metallic alloys improves with grain refinement and therefore Inconel 718 components are fabricated with wrought material with fine microstructures (grain sizes $\le $ ASTM 6 ) \citep{Alexandre2004}.  However, optimization of the fatigue behavior of the components requires a quantitative understanding of the relation between fatigue performance and grain size distribution and of the forming processes to tailor the grain size distribution in the component. This paper is focused in the first objective, the analysis and quantification of the effect of grain size on the LCF behavior of wrought Inconel 718.

The effect of the grain size in the LCF behavior of wrought Inconel 718 has been analyzed in many experimental investigations \citep{Pieraggi1994, Spath2001, Song2005, Alexandre2004, Bhowal2008}, which showed an increase in fatigue life as the grain size decreased. The effect of the grain size on the fatigue life in Inconel 718  depends on the applied cyclic strain range and it is much more noticeable at  small cyclic strain ranges \citep{Song2005}. In addition, the fatigue crack nucleation mechanisms in this alloy are different in the case of ultrafine grain sizes ($\le $ 10$\mu$m)  and fine and coarser grain sizes ($>$ 20 $\mu$m).  Crack initiation is controlled by the presence of second phase particles in the former, while cracks are nucleated along persistent slip bands and propagated through the grains in the latter \citep{Alexandre2004,Bhowal2008,Abikchi2013}. 

The mechanisms responsible for the effect of grain size on the fatigue life are not fully established and depend on the particular alloy, the range of grain sizes and the applied cyclic plastic strain amplitude. Two common approaches have been followed to analyze this problem from the simulation viewpoint. The first one is based on the dislocation-based model for fatigue crack initiation proposed by Tanaka and Mura \citep{Tanakka1981}. In this model, the grain size appears explicitly in the relationship between the number of cycles necessary for crack nucleation and the accumulated plastic strain (or any other fatigue indicator parameter chosen) in each slip system of each grain \citep{Shenoy2007, Castelluccio2015, Castelluccio2016}. In the particular case of Inconel 718, Alexandre {\it et al.} \cite{Alexandre2004}  proposed a macroscopic model for Inconel 718 in which the number of fatigue cycles for crack initiation was inversely proportional to the square of the cyclic plastic strain amplitude and of the grain size. The second approach assumes that the grain size effect is mainly due to the effect of grain boundaries which hinder the slip transfer between grains, leading to the formation of pile-ups and to hardening as the grain size decreases \citep{Hansen2004}. In this case, the constitutive equation (i.e., the crystal plasticity model) has to incorporate the effect of grain size, leading to changes in the accumulated plastic strain (or any other fatigue indicator parameter) as a function of the grain size \citep{Sweeney2014,Sweeney2014b,Wan2014}. 

Within the framework of computational homogenization using crystal plasticity, the incorporation of size effects in the constitutive equation can be achieved  by means of strain gradient plasticity models to simulate the formation of dislocation pile-ups at the grain boundaries \citep{Shu1999, Evers2002} and this strategy was followed by Sweeney {\it et al.} \cite{Sweeney2014}.  However, strain gradient crystal plasticity models are non-local: its implementation in standard finite element software presents some fundamental issues \cite{RodriguezGalan2017} and the models are usually  too expensive from the computational viewpoint to simulate  large Representative Volume Elements (RVEs) of the microstructure during several fatigue cycles. An alternative strategy to include the grain size in the constitutive equation was recently developed by Haouala {\it et al.} \cite{Haouala2017}, who used a Taylor model to relate the critical resolved shear stress in each slip system with the dislocation density. The evolution of the dislocation density during deformation took into account the dislocation storage at the grain boundaries by including the distance to the grain boundary to compute the dislocation mean free path. As a result, the critical resolved shear stress increased as the distance to the grain boundary decreased, and this dislocation-based model was able to reproduce the effect of grain size on the yield strength of  polycrystalline Cu under monotonic deformation. However, its extension to complex alloys under cyclic deformation is far from trivial. Alternatively, Shenoy {\it et al.} \citep{Shenoy2008} proposed a simpler approach in which the critical resolved shear stress in each grain depends on the grain size. This option can be considered a simplification of the model in \cite{Haouala2017}, because of the mechanical response of each grain is homogeneous throughout the grain but depends on the grain size, and it can be readily incorporated in  both phenomenological \citep{Sun2012} or physically-based crystal plasticity models \citep{Shenoy2008}.  The common choice to relate the critical resolver shear stress with the grain size is based on the empirical relations established by Hall \citep{Hall1951} and Petch \citep{Petch1953} for polycrystals.

The objective of this work is to model the effect of the grain size in the fatigue life of wrought Inconel 718 with coarse and fine grain sizes. The mechanical performance of the polycrystal was simulated by means of computational homogenization of an RVE of the microstructure. The single crystal behavior was introduced with a phenomenological crystal plasticity model. The effect of grain size on the critical resolved shear stress and on the back stress was included  through a Hall-Petch relationship that was calibrated to provide the appropriate mechanical response of the polycrystal under cyclic deformation for two different grain sizes. Crack nucleation was assumed to be controlled the formation of slip bands and fatigue indicator parameters were computed from the numerical simulations of the cyclic deformation of the RVEs. They were used to predict the fatigue life as a function 
of the applied cyclic strain range and the grain size and were in good agreement with the experimental results.

The paper is organized as follow. The LCF tests performed at $400^{\circ}$C and  {$R_\varepsilon$} = -1 for both coarse-grain (ASTM 3) and fine-grain (ASTM 8.5) wrought Inconel 718 are described. The numerical simulation of cyclic deformation by means of computational homogenization and the fatigue life prediction model are described in section 3, while the experimental results for the fatigue life and the model predictions are compared and discussed in section 4.  Finally,  the main conclusions of the work are found in section 5.

\section{Experimental methodology}
\subsection{Microstructure}

\begin{figure}[H]
\includegraphics[scale=0.25]{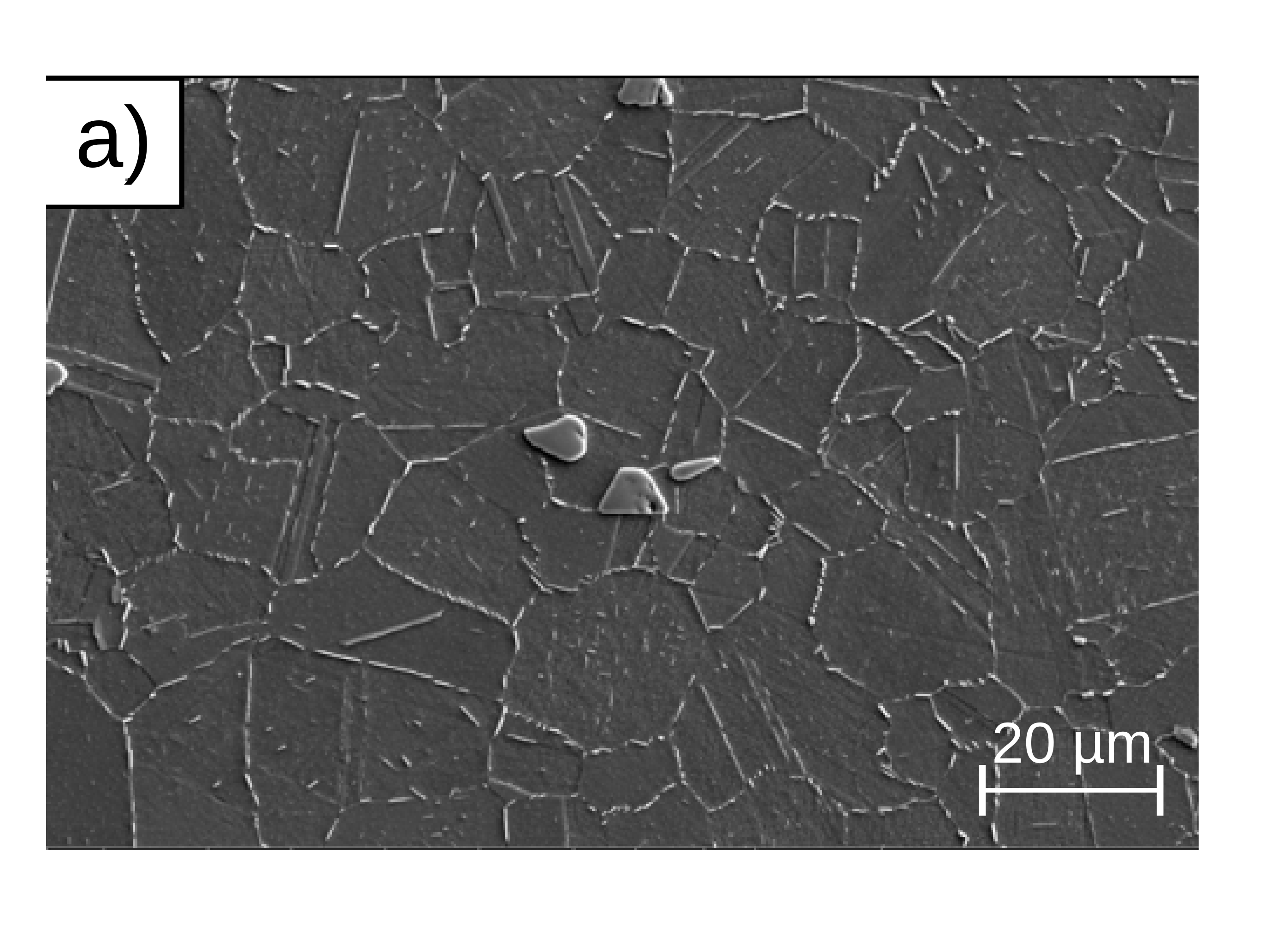}
\includegraphics[scale=0.25]{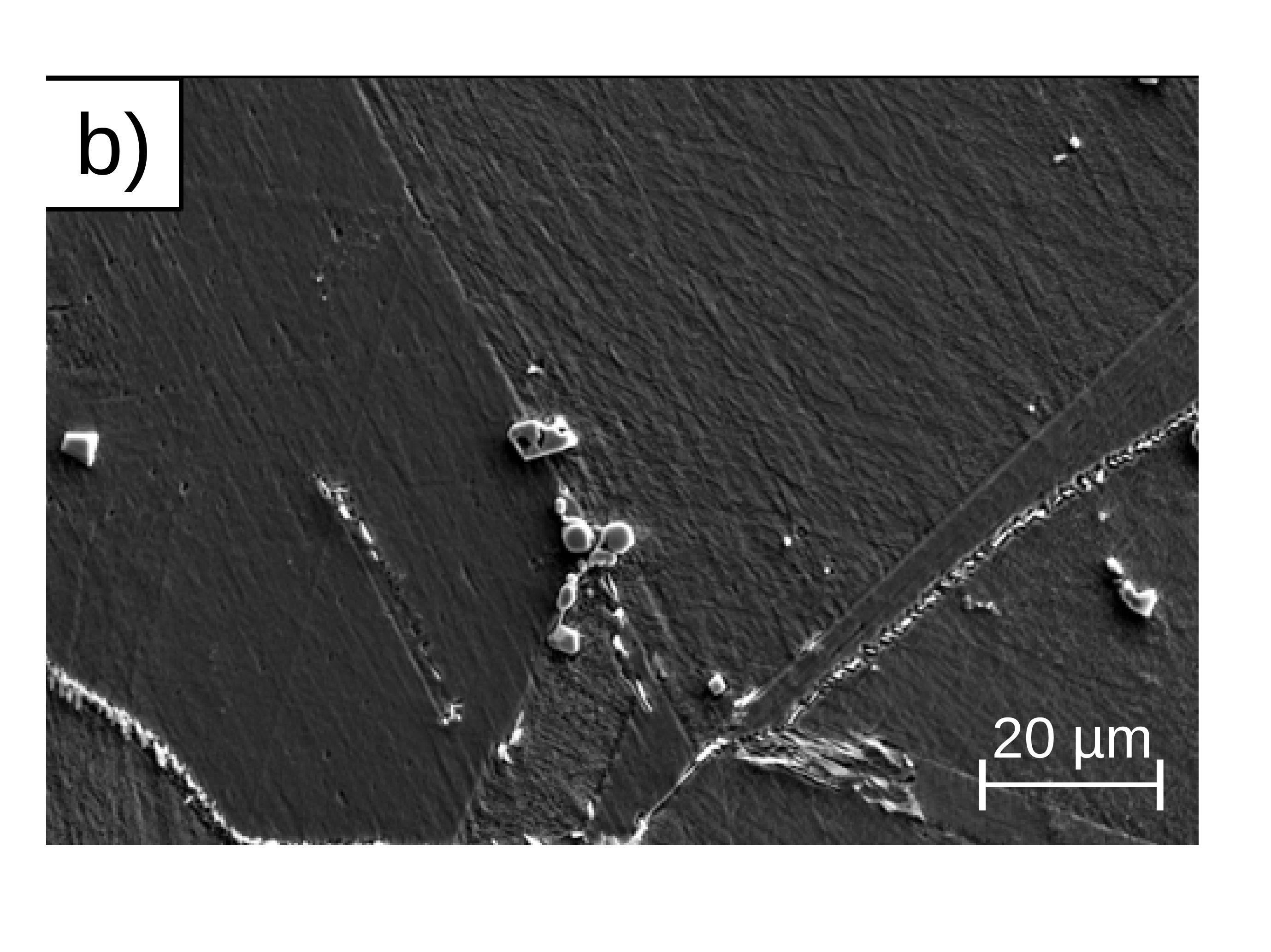}
\caption{Scanning electron micrographs of the microstructure of wrought IN718.  a) Fine grain size (ASTM 8.5). b) Coarse grain size (ASTM 3).} 
\label{microstructure}
\end{figure}

The fatigue behavior of wrought Inconel 718, named from here on IN718 alloy,  with two different grain sizes was measured. The average grain size of each material was obtained by the intercept method following the standard ASTM E112, leading to fine grain (ASTM grain size 8.5, $\approx$ 20 $\mu$m) and a coarse grain (ASTM grain size 3, $\approx$ 125 $\mu$m) microstructures, which are shown in Figs. \ref{microstructure}a) and b), respectively.  Both microstructures are mainly composed by a solid solution FCC $\gamma$ phase with negligible volume fractions of dispersed metal carbides ($<$ 1\%). The amount of $\delta$ phase was  $<$ 1\% in the coarse-grain material and in the {range of 10$\%$} in the fine-grain material. Transmission electron microscopy analysis, performed using a JEOL 3000F instrument, showed the presence of $\gamma$' and $\gamma$'' precipitates in the interior of the grains with sizes below 50 nm.

\subsection{Mechanical characterization}

The tensile stress-strain behavior of the two different wrought IN718 alloys was determined using an universal testing machine. Tests were carried out at 400$^{\circ}$C and a constant strain rate of 5.0 10$^{-3}$ s$^{-1}$. The tests were carried out on cylindrical smooth specimens with a diameter of 5.08 mm and  a gauge length of 12.7 mm for the ASTM 8.5 microstructure and a diameter of 6.35 mm and  a gauge length of 12.7 mm for the ASTM 3 alloy. 
The LCF tests at 400$^{\circ}$C of the two microstructures were performed on a MTS servo hydraulic fatigue load frame  with a 100 KN load cell, according to the standard ASTM E606-04. The axial displacement in the central zone of the specimen was measured with a MTS extensometer directly mounted on the gauge length. A trapezoidal wave form was applied according to: 1s (dwell) - 5 10$^{-3}$ s$^{-1}$ (ramp up) - 1s (dwell)- 5 10$^{-3}$ s$^{-1}$ (ramp down). The tests were carried with  {strain ranges} of  {$\Delta \varepsilon/\Delta \varepsilon_{min}$} = 1, 1.5, 2, 2.5, 3 and 3.5 under fully reversed cyclic deformation  $R_\varepsilon$ =  -1, where  $\Delta \varepsilon_{min}$ stands for  the minimum cyclic strain range applied in the fatigue tests.

\section{Computational Homogenization Strategy}

\subsection{Numerical model}

\begin{figure}[H]
\includegraphics[width=0.5\textwidth]{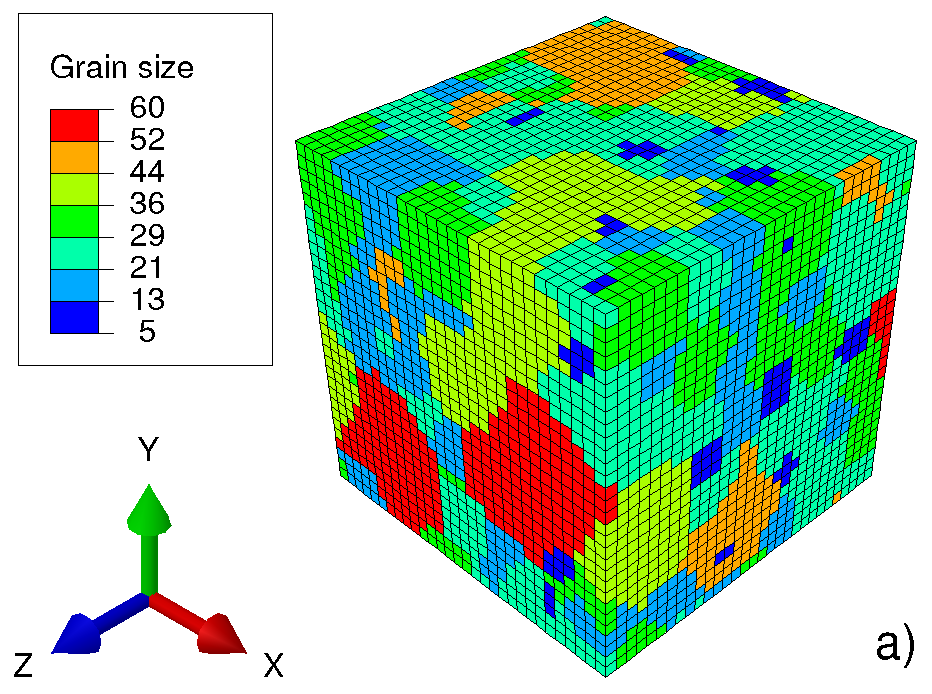}
\includegraphics[width=0.5\textwidth]{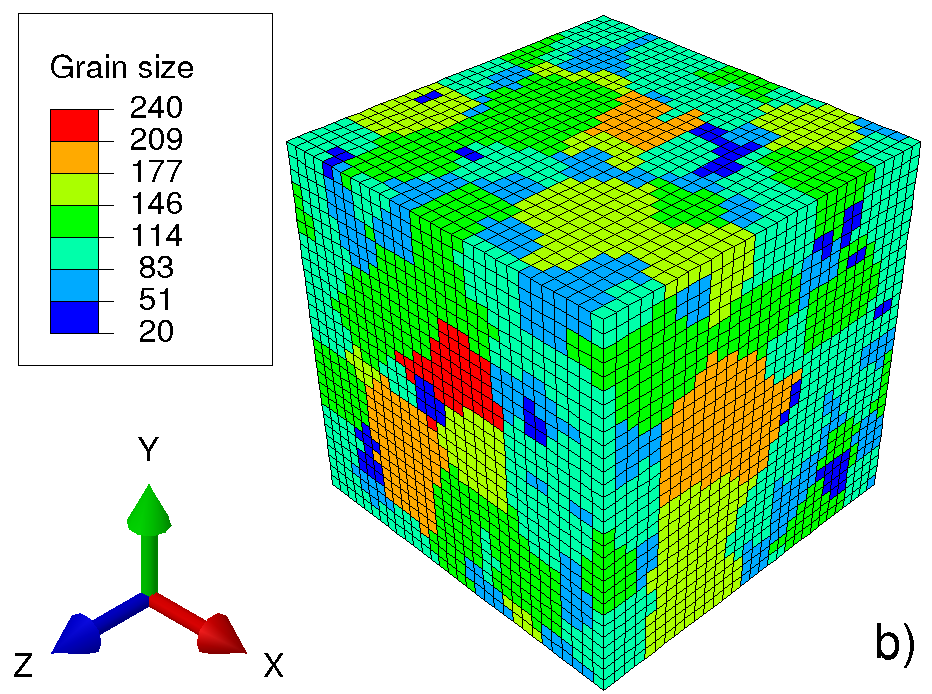}
\caption{Finite element model of the RVE of wrought polycrystalline {IN718 alloy}. a) Average grain size ASTM 8.5. b) Average grain size ASTM 3. The grain size in the legend corresponds to the equivalent diameter in $\mu$m.}  
\label{Polycrystal}
\end{figure}

The effective polycrystal behavior was obtained by computational homogenization using the finite element method. The microstructure was represented by a cubic RVE, which was discretized with a regular mesh of $N$ x $N$ x $N$ cubic C3D8 finite elements where each crystal is represented by many elements. 
The RVEs approximately include 300 grains and 90 elements per grain. It was demonstrated that this size was accurate enough of the statistical representation of the grain orientations and microfields in different grains \citep{Cruzado2017,Cruzado2017b}. The texture of the grains in the RVE was random, following the experimental data \citep{Cruzado2017}.
The grain size distribution in the RVE followed  the experimental grain size distribution, which was measured from a cross-section that include approximately 300 grains. The experimental 2D grain size distribution was transformed into a 3D distribution assuming spherical grains using StripStar \citep{HEILBRONNER1998}. The 3D grain radius distribution was approximated by a lognormal function whose average and standard deviation are given by $\sigma$ = 0.475 and $\mu$ = 3.5691 for the ASTM 3 alloy and  by $\sigma$ = 0.544 and $\mu$ = 2.031 for the ASTM 8.5 alloy. These parameters  were used in the software Dream3D \cite{Dream3D} to generate the RVE. Because the number of grains in each RVE was limited, 20 RVEs were generated for each material to take into account the statistical aspects of the grain size distribution based on previous results of fatigue simulations in {IN718 alloy} \cite{Cruzado2017}. The finite element models of two RVEs corresponding to the fine and coarse grain size distributions are depicted in Fig. \ref{Polycrystal}.

Numerical simulations were carried out using periodic boundary conditions in Abaqus/standard. The macroscopic loading conditions are introduced in the model by applying displacements and loads in the master nodes in order to achieve the target deformation history, see \cite{Segurado2013} for more details. The crystal behavior was introduced following the crystal plasticity model described in the next section. 

\subsection{Crystal plasticity model}

The elasto-plastic behaviour of {IN718 alloy} grains was based in the phenomenological crystal plasticity  model proposed by Cruzado {\it et al.} \cite{Cruzado2017}, which is briefly recalled here for the sake of completion.  The model was able to reproduce accurately the main cyclic characteristics of {IN718 alloy}, namely kinematic hardening, mean stress relaxation and cyclic softening.  

The crystals were assumed to behave as elasto-viscoplastic solids in which the plastic slip rate for a given slip  system follows a power law given by:

\begin{equation}
\label{eq_(7):gamma_dot}
\dot{\gamma}^\alpha=\dot{\gamma}_0 \left(
  \frac{| \tau^\alpha-\chi^\alpha |}{g^\alpha}\right)^\frac{1}{m}\mathrm{sign} (\tau^\alpha-\chi^\alpha)  
\end{equation}

\noindent where $\dot{\gamma}_0$ is the reference strain rate, ${m}$ the rate sensitivity parameter, ${g^\alpha}$ the critical resolved shear stress of the ${\alpha}$ slip system and $\chi^\alpha$ the back stress. 

Grain boundaries hinder the plastic slip and lead to the formation of dislocation pile-ups at grain boundaries, increasing the dislocation density in form of Geometrically Necessary Dislocations (GNDs). The amount of GNDs depends on the actual grain size, and gives rise to an effective increase in the flow stress of the crystal. This effect can be introduced either by assuming that the density of GND  mainly contributes to the isotropic hardening through the Taylor relation \cite{CHEONG20051797,MA20067287} or by considering that the development of GNDs mainly influences the kinematic hardening \cite{EVERS20045209} though the back stress evolution. The effect of grain size was included in both terms in this phenomenological approach.

It is well established that the GND density in polycrystals increase as the grain size decreases. This increase contributes to the isotropic hardening because these dislocations will act as additional obstacles to the movement of other dislocations, increasing the critical resolved shear stress,  $g^\alpha$, in eq. (\ref{eq_(7):gamma_dot}). This term includes two contributions that determine the amount of hardening or softening under monotonic ($g^\alpha_m$) and cyclic (${g}_c$) deformation according to 

\begin{equation}
{g}^\alpha=g^\alpha_m + {g}_c
\label{isotropic}
\end{equation}

\noindent  where ${g}^\alpha_m$ controls the evolution under monotonic deformation and ${g}_c$ determines the cyclic softening due to a progressive reduction of the critical resolved shear stress induced by changes in the direction of plastic shear. 

In the case of monotonic hardening, the effect of grain size is only introduced  in the initial value of the critical resolved shear stress, $\tau_0$, for  $g^\alpha_m$, which is given by

 \begin{equation}
\label{hall_petch}
\tau_0=\tau_0'+\frac{k_0}{\sqrt{D}}  
\end{equation}

\noindent where $\tau_0'$ corresponds to the initial critical resolved stress of a very large grain (in which grain boundary effects can be neglected), $D$ the grain diameter and $k_0$ the Hall-Petch constant. 

The evolution of $g^\alpha_m$ with the applied strain followed the crystal plasticity model of Cruzado {\it et al.} \cite{Cruzado2017} and  is obtained from the contribution of the shear strain of all the slip systems $\beta$ in the crystal according to

\begin{equation}
\label{eq_(10):tau_crit}
\dot{g}^\alpha_m=\sum_\beta h |\dot{\gamma}^\beta|   
\end{equation}

\noindent  where the self hardening modulus $h$, negative in the case of IN718 to account for the initial softening \cite{Cruzado2017}, follows the Asaro-Needleman isotropic hardening model according to,

\begin{equation}
\label{eq_(12):Assaro}
 h\left(\gamma_a\right)=-h_0 \mathrm{sech}^2 \left| \frac{h_0\gamma_a}{\tau_s-\tau_0}\right| 
\end{equation}

\noindent where $\tau_0$ is the initial critical resolved shear stress,
 $h_0$ the initial softening modulus, $\tau_s$ the saturation stress (which is smaller than $\tau_0$) and  $\gamma_a$ the accumulated shear strain in all slip systems of the crystal, which is obtained as 

\begin{equation}
\label{eq_(13):gamma_acum}
\gamma_a=  \sum_\alpha \int_0^t \!  |\dot{\gamma}^\alpha|  \, \mathrm{d}t.
\end{equation}

Thus, the contribution of grain size to the isotropic hardening/softening (eq. \ref{eq_(12):Assaro}) during monotonic deformation has been neglected for simplicity because this term in very small as compared to the kinematic hardening in this alloy.

The evolution of flow stress due to the cyclic deformation, ${g}_c$, leads to cyclic softening in {IN718 alloy} and it was given by a Voce type law with negative slope, according to \cite{Cruzado2017},

\begin{equation}
\label{eq_(14):cyclic_softening}
g_c=-\left(\tau_s^{cyc}+h_2\gamma_{cyc}\right)\left(1-\exp^\frac{-h_1\gamma_{cyc}}{\tau_s^{cyc}}\right)
\end{equation}

\noindent where $\tau_s^{cyc}$ is the saturation softening (the maximum reduction of the critical resolved shear stress due to cyclic softening), $h_{1}$  and $h_2$ the cyclic softening parameters, and $\gamma_{cyc}$ (the cyclic accumulated plastic strain) is an internal variable  to capture the cyclic softening under a general loading history. $\gamma_{cyc}$ is given by

\begin{equation}
\label{eq_(15):cyclic_plast}
\gamma_{cyc}=\sum_\alpha \int_0^t |\dot{\gamma}^\alpha | \mathrm{d}t - \sum_\alpha \left|\int_0^t \dot{\gamma}^\alpha \mathrm{d}t\right|.
\end{equation}

The effect of grain size was not introduced in ${g}_c$  because this phenomenon is related with the progressive shearing of the precipitates and only depends on the amount of $\gamma''$ precipitates, that is constant for all grain sizes. 

Dislocation pile-ups at the grain boundaries provide the main contribution to the kinematic hardening at the crystal level because the back stresses will assist dislocation motion during reverse deformation. They induce a size effect because the density of GNDs increases as the grain size decreases. Thus,  the kinematic hardening $\chi^\alpha$ is expressed as, 

\begin{equation}
\label{eq_(9):kinematic_OWM}
\dot{\chi}^\alpha=c\dot{\gamma}^\alpha-d\chi^\alpha|\dot{\gamma}^\alpha|\left( \frac{|\chi^\alpha|}{c/d}\right)^{mk}   
\end{equation}

\noindent where $mk$ is the parameter that controls the mean stress relaxation velocity. $c$ stands for the direct hardening modulus and the ratio $c/d$ for the saturation hardening value. They are given by

\begin{equation}
\label{eq_(9):kinematic_hall_c}
c=c'+\frac{k_c}{\sqrt{D}}
\end{equation}

\begin{equation}
\label{eq_(9):kinematic_hall_d}
\frac{c}{d}=\left(\frac{c}{d}\right)'+\frac{k_{c/d}}{\sqrt{D}}
\end{equation}

\noindent where  $c'$ and $(c/d)'$ stand for the direct hardening modulus and saturation hardening, respectively, corresponding to a very large grain (where no dislocation pile-ups are formed) and $k_c$ and $k_{c/d}$ stand for the Hall-Petch parameters that introduce the effect of grain size.

The model parameters for {IN718 alloy} at 400$^{\circ}$C are presented in Table \ref{Table:CP_parameters} (note that the viscoplastic parameters are normalized by $\tau_0'$).  The elastic constants were obtained from the experimental values reported in \citep{Martin2014} at room temperature assuming a linear reduction of the three elastic constants with temperature similar to the one found for the elastic modulus of IN718 polycrystals. The reference strain rate, $\dot{\gamma}_0$, and the strain rate sensitivity exponent, $m$, were obtained from mechanical tests on single crystal micropillars machined from the polycrystals \citep{Cruzado2015}. The parameters defining cyclic softening  and the mean stress relaxation, unaffected by the grain size, were directly obtained from \cite{Cruzado2017}. The initial critical resolved shear stress which included the Hall-Petch constant was obtained from the monotonic uniaxial tensile tests of both microstructures. The remaining parameters of the crystal plasticity model,  that introduce the grain size effect in the kinematic hardening, were obtained  from the experimental results of the cyclic response of the fine and coarse grain size IN718 at 400$^{\circ}$C with  $R_\varepsilon$ = -1  and $\Delta \varepsilon/\Delta \varepsilon_{min}$=3.

\begin{table}[H]
\centering
\begin{tabular}[5pt]{lccccc}
\hline\noalign{\vskip2pt}
\multirow{2}{*}{Elastic}
  &C$_{11}$(GPa)  &   C$_{12}$(GPa)   &   C$_{44}$(GPa)    \\[0.2cm]
& 240 & 165 & 101  \\
\hline\noalign{\vskip2pt}
\multirow{2}{*}{Viscoplastic}
 &  $m$ & $\dot{\gamma}_0$ &  &   &   \\[0.2cm]
 & 0.017 &    2.42 $10^{-3}$   &      &   &       \\[0.5cm]
 \hline\noalign{\vskip2pt}
 \multirow{2}{*}{Isotropic softening} & $\tau_0'$ (MPa)  & $\tau_s$ & $h_0$  &  $k_0$  \\[0.2cm]
& $\tau_0'$ & 0.83$\tau_0'$ & 57.29$\tau_0'$ &  8.14 $10^{-4}$  $\tau_0'$ \\[0.5cm]
\hline\noalign{\vskip2pt}
\multirow{2}{*}{Kinematic hardening}  & $c'$ & $(c/d)'$ & $mk$ & $k_c$ & $k_{c/d}$ \\[0.2cm]
& 33.8$\tau_0'$ & 0.084$\tau_0'$  & 20.49 & 1.02 $10^{-3}$ $\tau_0'$ & 0.145$\tau_0'$ \\[0.5cm]
\hline\noalign{\vskip2pt}
\multirow{2}{*}{Cyclic softening} & $\tau_s^{cyc}$ & $h_{1}$ & $h_2$ &   \\[0.2cm]
  & 0.083$\tau_0'$ & 0.072$\tau_0'$ & 3.44 $10^{-6}$  $\tau_0'$ &\\[0.5cm]
\noalign{\vskip2pt}\hline
\end{tabular}
\caption{Parameters of the crystal plasticity model for cyclic deformation of wrought IN718 alloy at $400^{\circ}$C including the effect of grain size.}
\label{Table:CP_parameters}
\end{table}

\subsection{Fatigue crack initiation model}

Prediction of fatigue crack nucleation by means of computational homogenization is carried out using  the local Fatigue Indicator Parameters  (FIP) \citep{Manonukul2004, Sweeney2015, Wan2014} which provide information about the evolution of the mechanical fields in each fatigue cycle. In the case of {IN718 alloy},  Cruzado {\it et al.} \citep{Cruzado2017b}  were able to predict the number of fatigue cycles for crack nucleation using the local crystallographic strain energy dissipated per cycle, $W_{cyc}^\alpha(\mathbf{x}$), which is given by

\begin{equation}
W_{cyc}^\alpha(\mathbf{x})= \int_{cyc} \tau_\alpha(\mathbf{x})\dot{\gamma}_\alpha(\mathbf{x}) \mathrm{d}t
\label{eq_(4):W}
\end{equation}

\noindent at the point $\mathbf{x}$ of the crystal, where $\tau_\alpha$ and $\dot{\gamma}_\alpha$ are the resolved shear stress and the shear strain rate on the slip system $\alpha$, respectively. The local value of the FIP defined in eq. (\ref{eq_(4):W}) is calculated at the centroid of each element in the RVE. The formation of a persistent slip band in which a crack will be incubated is a non-local phenomena that affects a finite volume of the grain. In order to consider this non locality in the model, the FIP values at the element centroids were averaged along bands parallel to the slip planes, as proposed by Castelluccio {\it et al.} \citep{Castelluccio2015}. This averaging has also a positive effect in reducing mesh dependency  as well as the spurious stress concentrations associated with the structured mesh. Under these assumptions, the FIP representative of a RVE subjected to cyclic deformation, $W_{cyc}^b$, is obtained as the maximum of the band-averaged local FIP throughout the RVE, according to:

\begin{eqnarray}
W_{cyc}^b=\max_{i=1,nb} \left \{\max_{\beta_i} \frac{1}{V_i}\int_{V_i} W_{cyc}^{\beta_i}(\mathbf{x})\mathrm{d}V_i \right \}
\label{eq_(5):FIPfinal}
\end{eqnarray}

\noindent  where $\beta_i$ (= 1, 2, 3) corresponds to the three different slips systems contained in the slip plane parallel to the band $i$, $V_i$ is the volume of that band and $nb$ the total number of bands in the microstructure, which is 4 times the number of elements in the RVE.

Under these conditions,  Cruzado {\it et al.} \citep{Cruzado2017b} showed that the number of cycles to initiate  crack, $N_i$, could be expressed in {IN718 alloy} by

\begin{equation}
N_i= \frac{W_{crit}^{NL}} {(W_{cyc}^b)^m}
\label{eq_(11):energy}                    
\end{equation}

\noindent where the fatigue life depends on the  cyclic FIP, $W_{cyc}^b$, through two material parameters, $W_{crit}^{NL}$ and $m$.

\section{Results and discussion} 

\subsection{Experimental results}
\begin{figure}[H]
\centering
\includegraphics[scale=0.82]{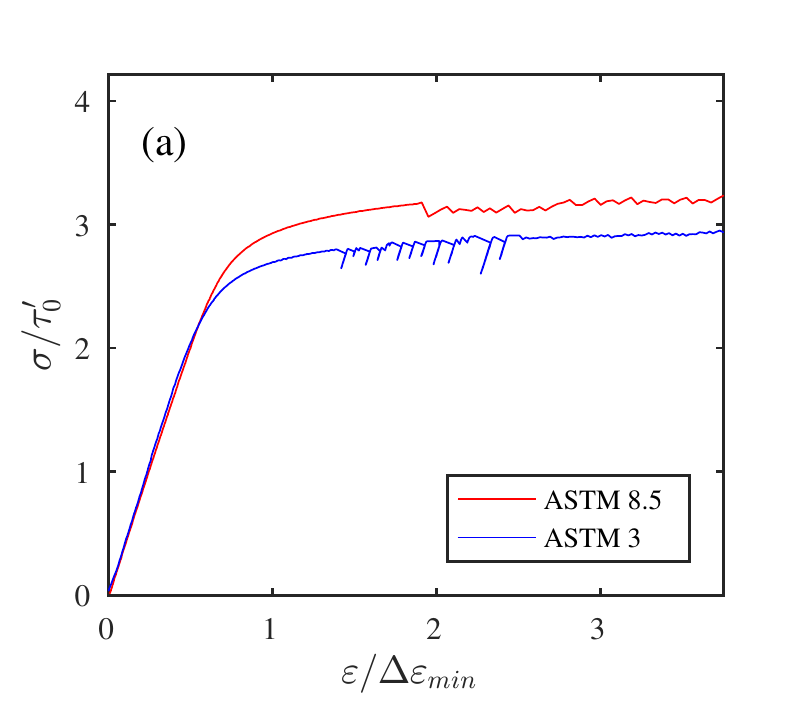}
\includegraphics[scale=0.82]{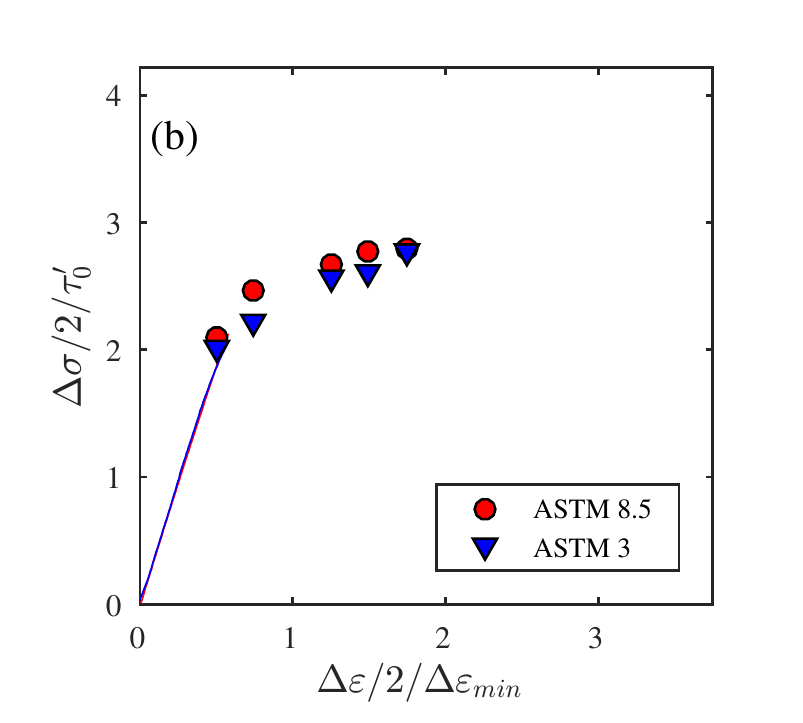}
\caption{Effect of the average grain size in the mechanical response of wrought {IN718 alloy} at 400$^{\circ}$C. a)
Tensile stress-strain curve. b) Stabilized cyclic stress-strain curve. The stresses are normalized by $\tau_0'$ and the strains by  {$\Delta \varepsilon_{min}$}, the minimum cyclic strain  {range} used in the fatigue tests.} 
\label{tensile_cyclic}
\end{figure}

The tensile and the stabilized cyclic stress-strain curves  at 400$^\circ$ are plotted in Fig.\ref{tensile_cyclic} for both IN718 alloys with different grain size. The tensile yield stress of the alloy increased by $\approx$ 10\% as the grain size decreased from ASTM 3 to ASTM 8.5, and the material with fine microstructure was also slightly stronger under cyclic deformation. It should be noted that the flow stress in the cyclic stress-strain curve was lower than that of the monotonic curve as a consequence of the characteristic cyclic softening of IN718 alloy.

\begin{figure}[H]
\includegraphics[scale=0.90]{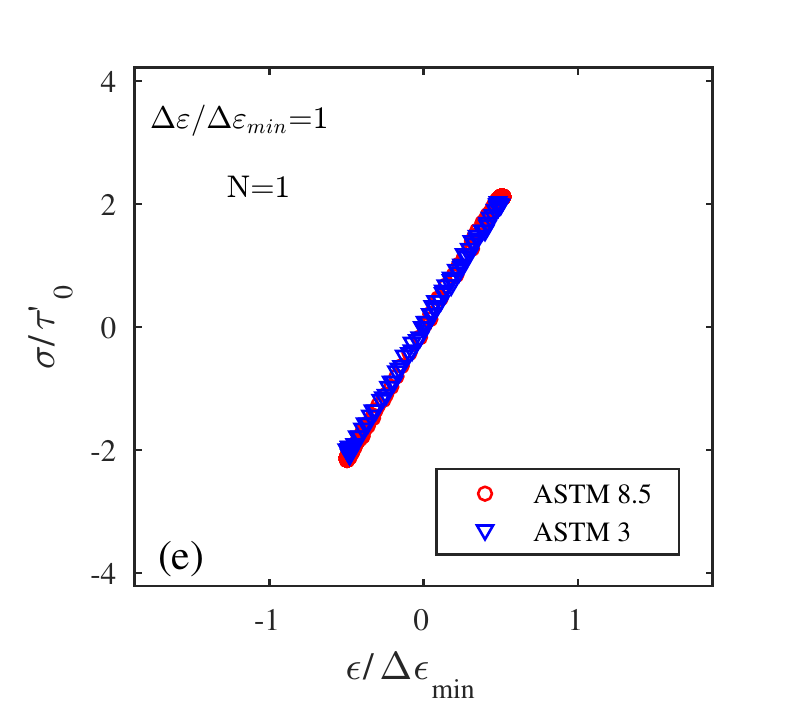}
\includegraphics[scale=0.90]{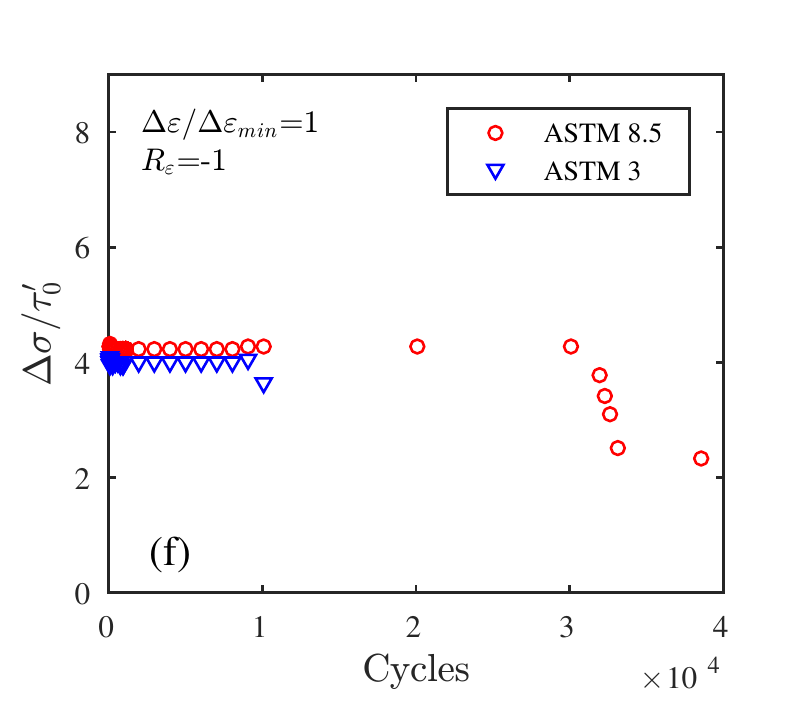}

\caption{Experimental results of the cyclic response of {IN718 alloy} at 400$^{\circ}$C with different average grain size tested under uniaxial cyclic deformation with  $R_\varepsilon$ = -1. (a) Cyclic stress-strain loops for  $\Delta \varepsilon/\Delta \varepsilon_{min}$=3. (b) Evolution of the stress range, $\Delta \sigma/\tau'_0$, with the number of cycles for  $\Delta \varepsilon/\Delta \varepsilon_{min}$=3. (c) {\it Idem} as (a) for  {$\Delta \varepsilon/\Delta \varepsilon_{min}$}=1.5. (d) {\it Idem} as (b) for  {$\Delta \varepsilon/\Delta \varepsilon_{min}$}= 1.5. (e) {\it Idem} as (a) for  {$\Delta \varepsilon/\Delta \varepsilon_{min}$}= 1. (f) {\it Idem} as (b) for  {$\Delta \varepsilon/\Delta \varepsilon_{min}$}=1.} 
\label{Cyclic_curves}
\end{figure}

The cyclic stress-strain behavior of IN718 alloy at 400$^{\circ}$C for three different values of the applied cyclic strain range (small, medium and large) is shown in Fig. \ref{Cyclic_curves}. The first loading cycle is represented in  Figs. \ref{Cyclic_curves}a), c), and e), while the evolution of the cyclic stress range as a function of the number of cycles is plotted in Figs. \ref{Cyclic_curves} b), d), and f).  These curves show that the stress range in the fine microstructure was always higher than that of the coarse grain material due to the strengthening provided by the smaller grain size. Moreover, the evolution of stress range with the number of cycles (cyclic softening) was the same for both microstructures. This behavior supports the hypothesis in the model which assumed that cyclic softening was independent of the grain size because this mechanism only depends on the $\gamma$'' precipitate distribution, which was equivalent in both microstructures.

\begin{figure}[H]
\includegraphics[scale=0.90]{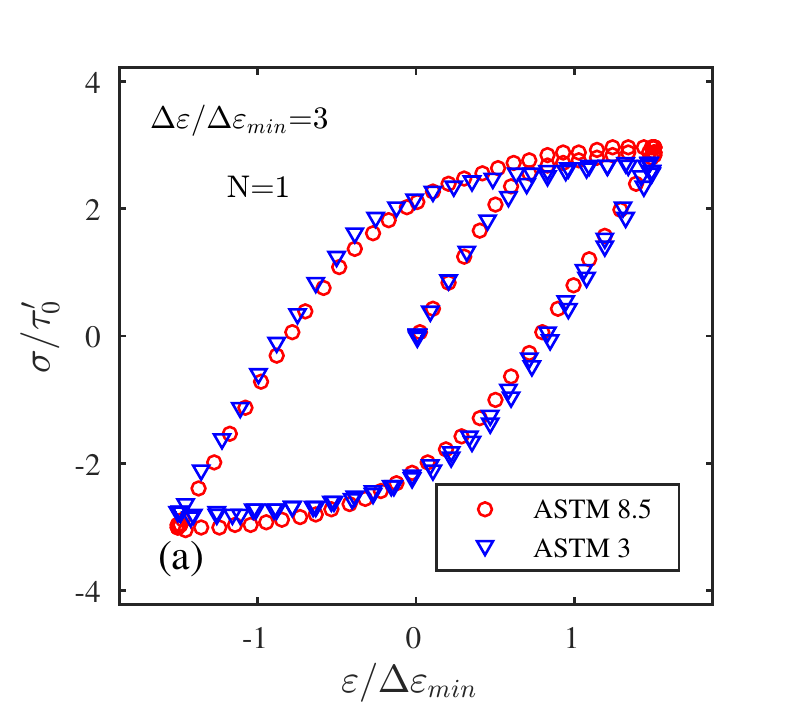}
\includegraphics[scale=0.90]{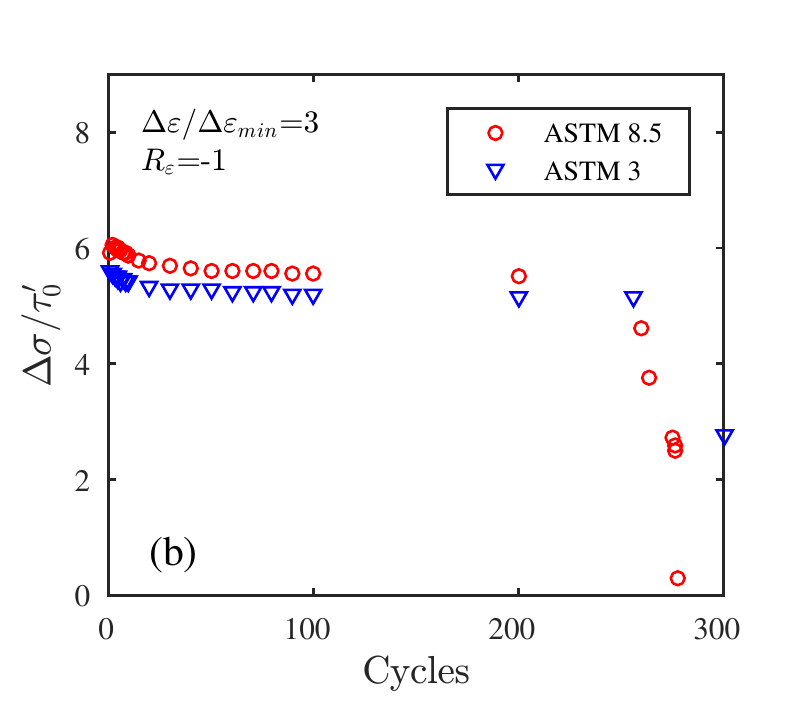}
\includegraphics[scale=0.90]{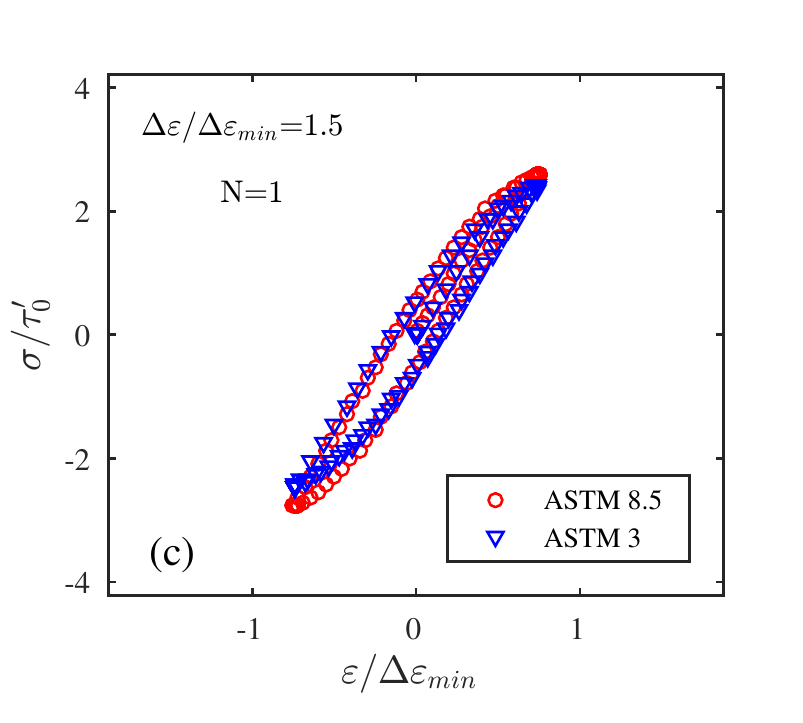}
\includegraphics[scale=0.90]{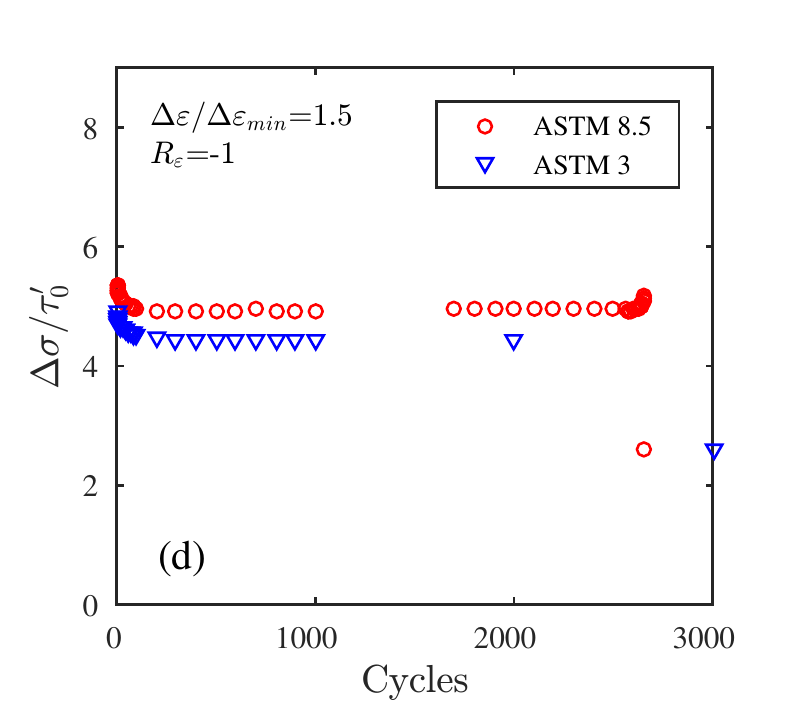}
\end{figure}

\begin{figure}[H]
\centering
\includegraphics[scale=0.90]{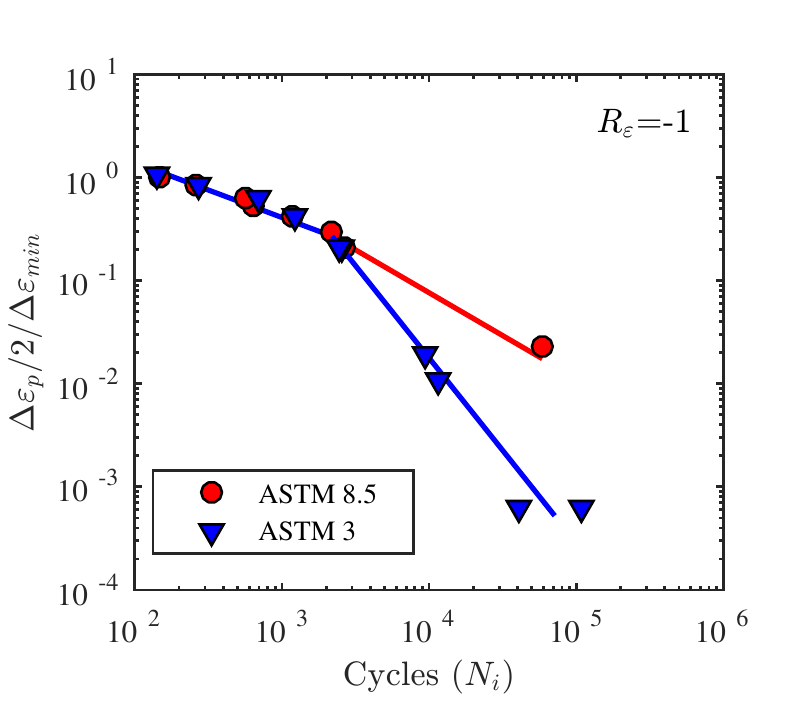}
\caption{Cyclic plastic strain amplitude,  $\Delta\epsilon_p/2$ (normalized by  {$\Delta \varepsilon_{min}$}) {\it vs.} the number of cycles to crack initiation, $N_i$, in {IN718 alloy} under fully-reversed cyclic deformation,  $R_\varepsilon$ = -1 at 400$^{\circ}$C.} 
\label{fatigue}
\end{figure}

The experimental results of the the cyclic plastic strain range (normalized by  $\Delta \varepsilon_{min}$) are plotted  in Fig.\ref{fatigue} {\it vs.} the number of cycles until crack nucleation, $N_i$, which was defined as the cycle in which the maximum load dropped by 5\%.
The dual slope Coffin-Manson behavior of IN718 alloy is shown in the figure where it can be observed that the transition between the two Coffin-Manson regimes (large and small cyclic strain ranges) took place at around 2000 cycles in both two microstructures. The fatigue life of both materials was equivalent for large cyclic plastic strain ranges and the differences in the strength observed in the cyclic stress-strain curves (Figs. \ref{Cyclic_curves}a and b) do not affect the fatigue life. On the contrary, the fine grain material exhibited much higher fatigue life when the applied cyclic plastic strain amplitudes were small and the hysteresis loops were nearly elastic (Fig. \ref{Cyclic_curves}b) and d). This fact indicates that the effect of grain size in the fatigue life of IN718 alloy only appears when plastic deformation is highly localized in very few grains  throughout the microstructure in which the plastic strains are slowly increasing during a significant fraction of the fatigue life. This failure is typical of IN718 alloy deformed at low cyclic plastic strain amplitudes \cite{Cruzado2017b}.

\subsection{Numerical results}

\begin{figure}[H]
\centering
\includegraphics[scale=0.9]{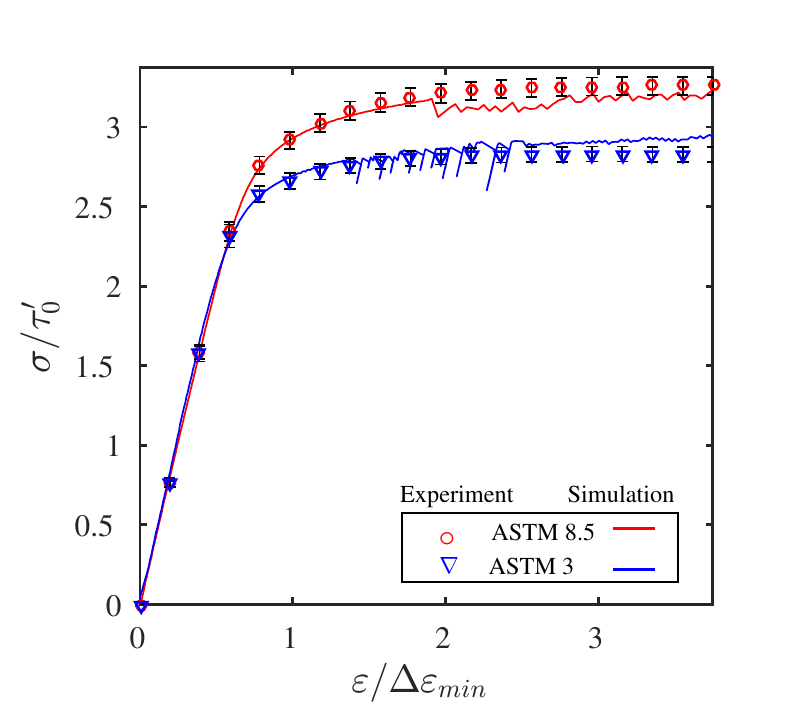}
\caption{Tensile stress-strain curved of wrought {IN718 alloy} at 400$^{\circ}$C. Stresses are normalized by $\tau_0'$ and strain by $\Delta \varepsilon_{min}$.} 
\label{monotonic_tensile}
\end{figure}

The crystal plasticity model presented above was applied to predict the monotonic tensile deformation of IN718 alloy with fine and coarse grain size. 20 RVEs were generated for each material using their grain size distributions and random texture. The resulting tensile stress-strain curves, obtained as the average of the 20 RVEs, are plotted in Fig. \ref{monotonic_tensile}. The error bars stand for the standard deviation obtained from 20 realizations for each material. The simulation results were in excellent agreement the experimental response for both grain sizes and it should be noticed that the parameters that include the grain size effect on isotropic and kinematic hardening were obtained from cyclic experiments. It is also worth noting that  the  standard deviation obtained in the simulations of the different realizations was small for both microstructures. This result indicates  that the size of each RVE was large enough to account for the effective behavior of the polycrystal.
 
\begin{figure}[H]
\includegraphics[scale=0.90]{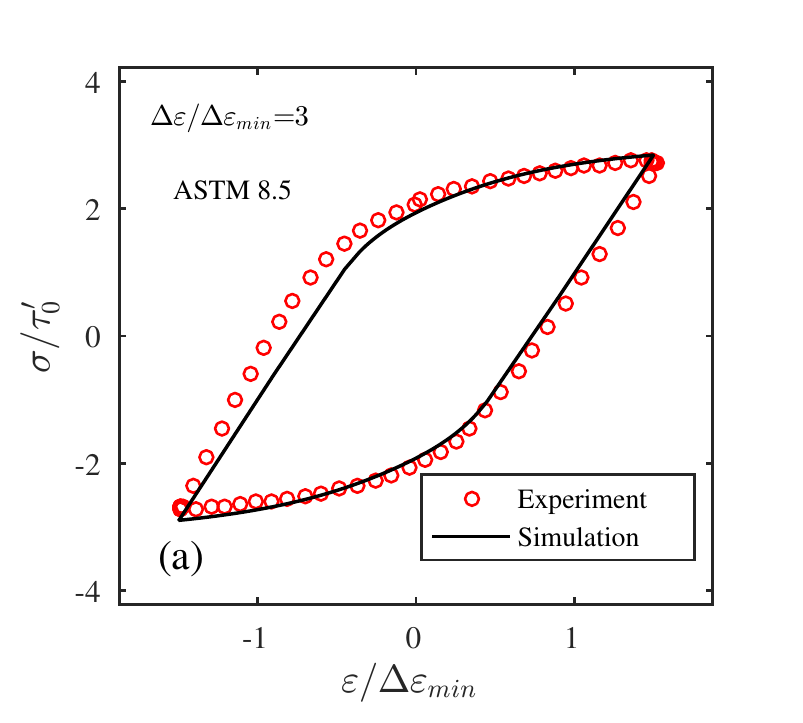}
\includegraphics[scale=0.90]{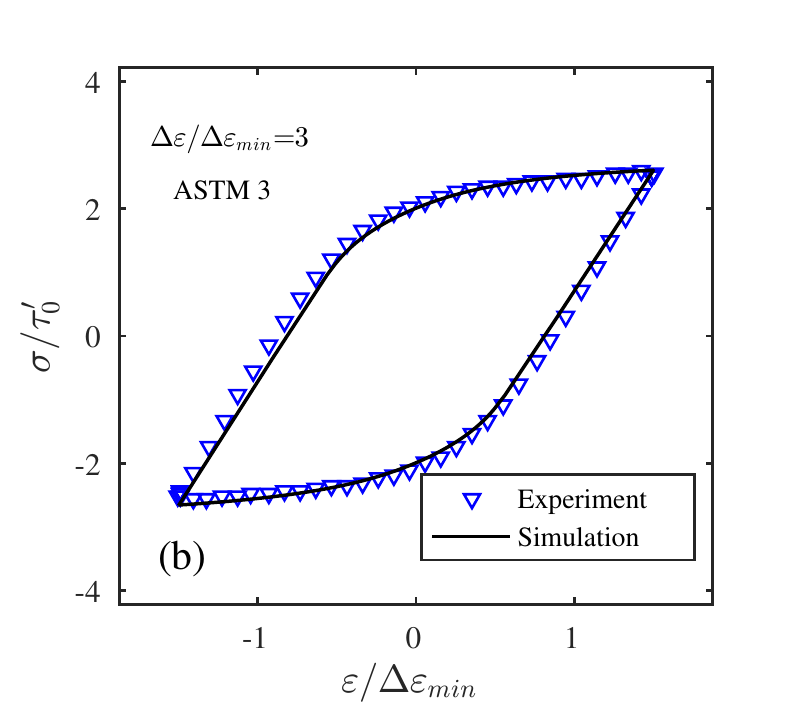}
\includegraphics[scale=0.90]{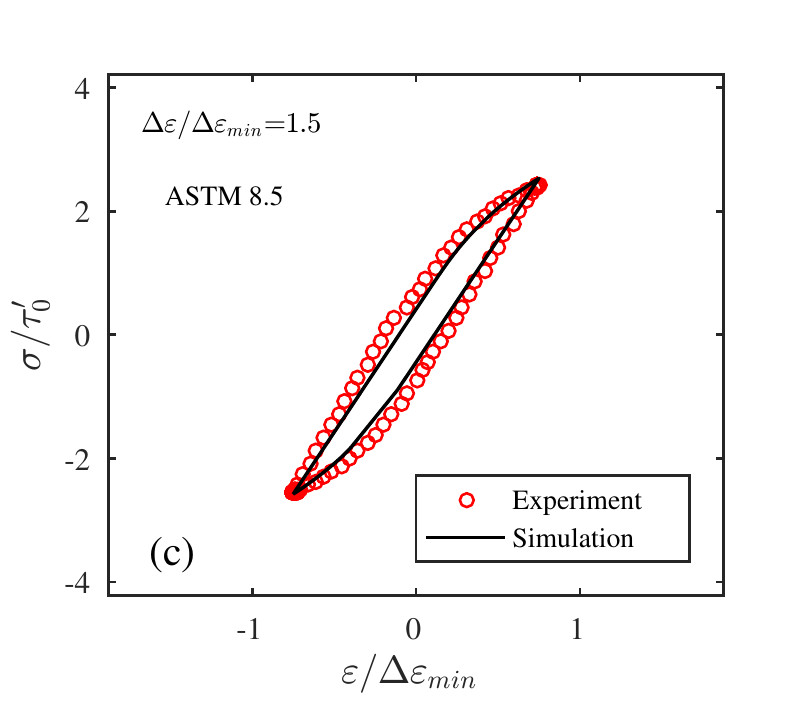}
\includegraphics[scale=0.90]{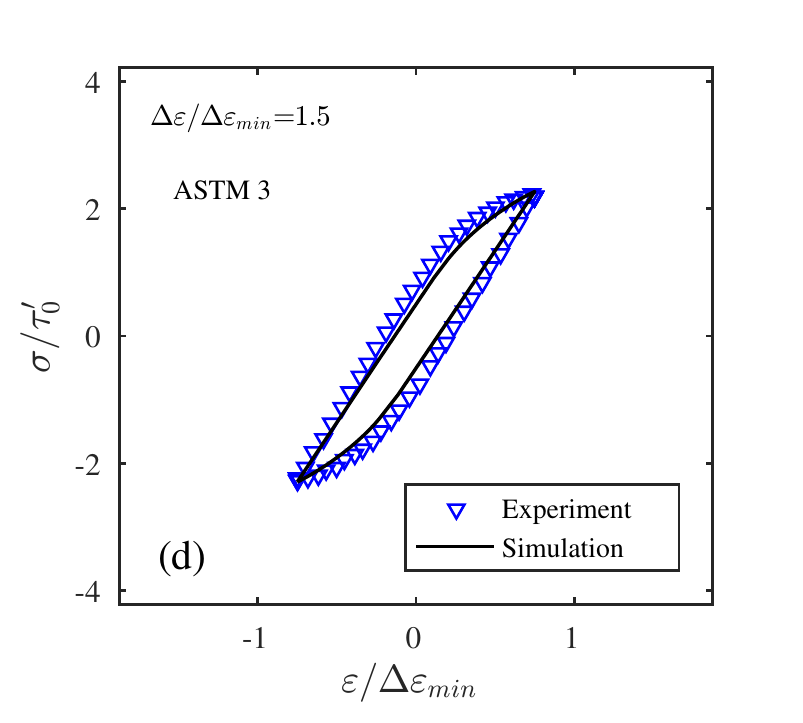}
\caption{Experimental results and numerical predictions of the stabilized cyclic stress-strain loops of {IN718 alloy} at 400$^{\circ}$C tested under uniaxial cyclic deformation with  $R_\varepsilon$=-1. (a)  {$\Delta \varepsilon/\Delta \varepsilon_{min}$} = 3 for ASTM 8.5 grain size. (b) {\it Idem} as (a) for ASTM 3 grain size. (c)  $\Delta \varepsilon/\Delta \varepsilon_{min}$= 1.5 for ASTM 3 grain size. (d) {\it Idem} as (c) for ASTM 3 grain size} 
\label{Cyclic_FEM}
\end{figure}

The cyclic strain-stress curves for both microstructures are plotted in Fig. \ref{Cyclic_FEM}, together with the experimental results, for two different applied cyclic strain ranges, namely  $\Delta \varepsilon/\Delta \varepsilon_{min}$ = 3 and 1.5. The numerical curves corresponds to the RVE whose behavior was the closest to the average monotonic stress-strain curves in Fig. \ref{monotonic_tensile} for each material. The stabilized stress-strain hysteresis loops measured in the experiments and predicted by the model for  both strain ranges were in good agreement. This result supports the phenomenological strategy used in the crystal plasticity model to account for the grain size effect {\it via} the introduction of Hall-Petch type relations in both the initial critical resolved shear stress and the back stress evolution. 

Finally, the parameters $W_{crit}^{NL}$ and $m$ in eq. (\ref{eq_(11):energy}), which are necessary to obtain the number of cycles for crack nucleation from the FIP, were obtained from the simulation of the cyclic behavior of two different tests ($\Delta \varepsilon/\Delta \varepsilon_{min}$ = 1 and 3.5) of the alloy with small grain size (ASTM 8.5) and the corresponding experimental results. It should be noted that the FIP, $W_{cyc}^b$, was obtained from the simulation of 20 different realizations of the RVE for each strain range. This is necessary because the number of grains in the RVE is very small as compared with that in the actual fatigue specimens and a minimum number of grains to obtain statistically-representative results can only be obtained from many realizations. The parameters of the fatigue life model were $W_{crit}^{NL}$ = 4.8485 x 10$^4$ MJ/m$^3$ and $m$ = 1.4755.

\begin{figure}[H]
\includegraphics[scale=0.90]{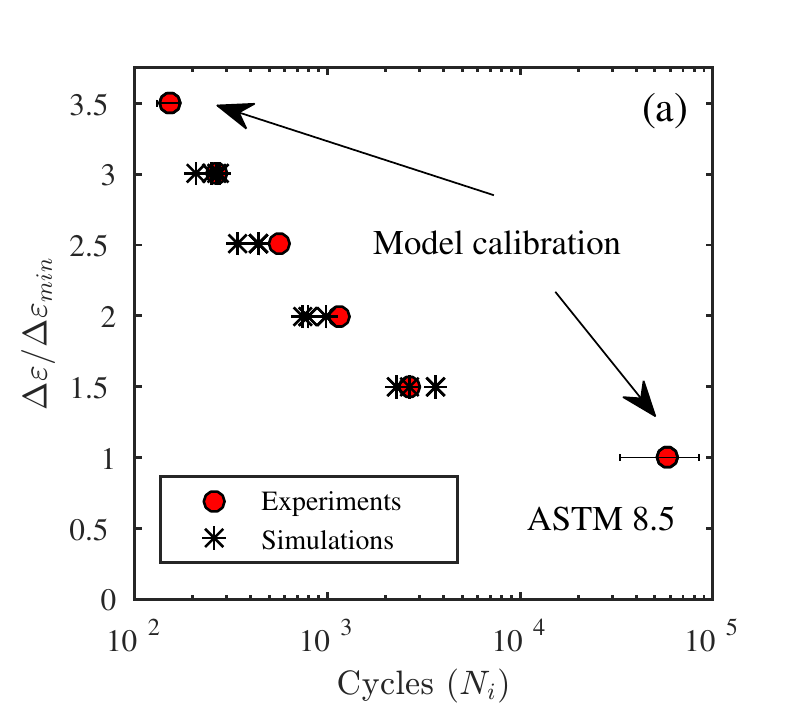}
\includegraphics[scale=0.90]{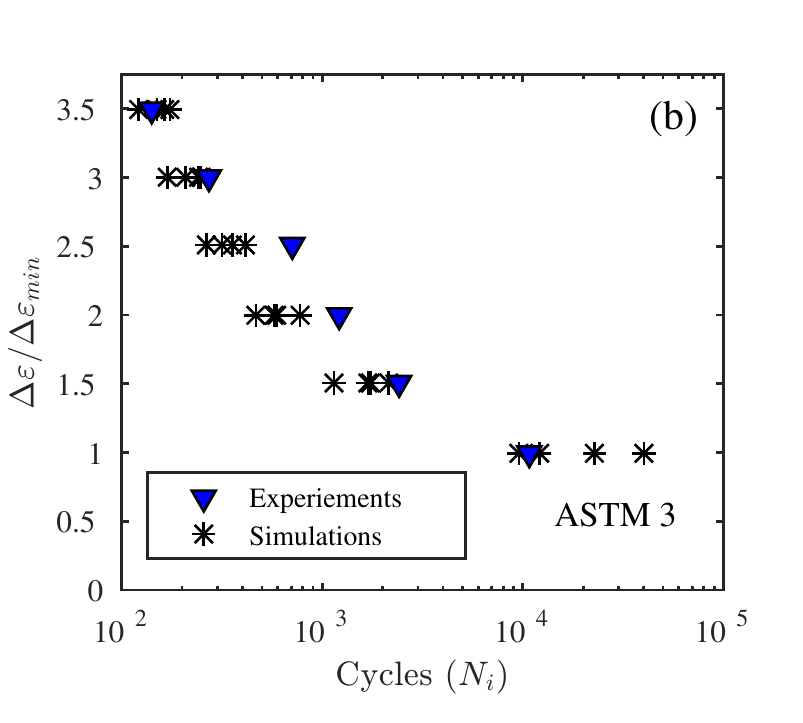}
\caption{ Experimental results and  model predictions for the number of cycles for fatigue crack initiation in {IN718 alloy} at 400$^{\circ}$C as a function of the applied cyclic strain range,  $\Delta \varepsilon$ (normalized by the minimum cyclic strain range,  $\Delta \varepsilon_{min}$). a) ASTM 8.5 grain size. b) ASTM 3 grain size. The results for 4 realizations of the RVE are shown in the case of the numerical simulations.}
\label{Fatigue prediction}
\end{figure}

The predictions of the number of cycles for fatigue crack initiation at 400$^\circ$C, $N_i$, as a function of the applied cyclic strain range,  $\Delta \varepsilon/\Delta\varepsilon_{min}$, are plotted in Figs. \ref{Fatigue prediction}(a) and (b) for each microstructure together with the corresponding experimental results. The two experiments used to calibrate the parameters of the fatigue life are marked in Fig. \ref{Fatigue prediction}(a) and the simulations results of these cases include the scatter bars corresponding to the simulations of  20 RVEs. For the remaining strain ranges considered, the numerical predictions correspond to a sub-sample of four different RVEs out of the 20 used to calibrate the fatigue model. 

The agreement between experiments and simulations was excellent in most cases for both grain sizes. The largest differences were found in the intermediate strain ranges for the coarse grain size where $N_i$ was slightly underpredicted. Moreover, the numerical simulations captured the differences in the fatigue behavior of IN718 alloy when subjected to large and small cyclic strain ranges (dual slope Manson-Coffin relationship). At large cyclic strain ranges, plastic deformation is homogeneously distributed throughout the microstructure and the influence of the details of the microstructure (including the average grain size) on the fatigue life is minimum. However, plasticity is localized in a few grains at low cyclic strain ranges and the different slope of the Manson-Coffin plot for a given microstructure reflects the change from homogeneous plastic deformation at large cyclic strain ranges to localized deformation at small cyclic  strain ranges.  

\begin{figure}[H]
\includegraphics[scale=0.90]{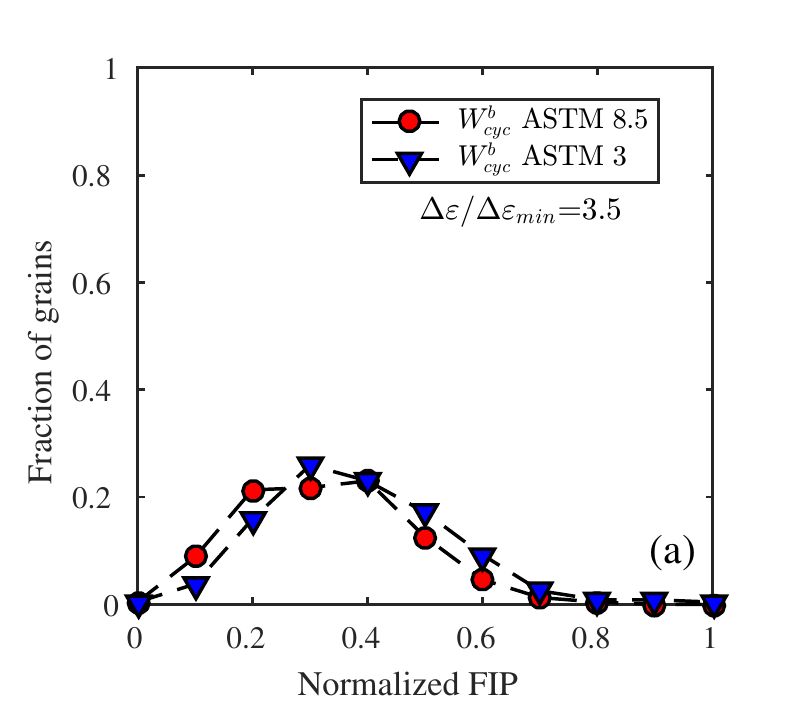}
\includegraphics[scale=0.90]{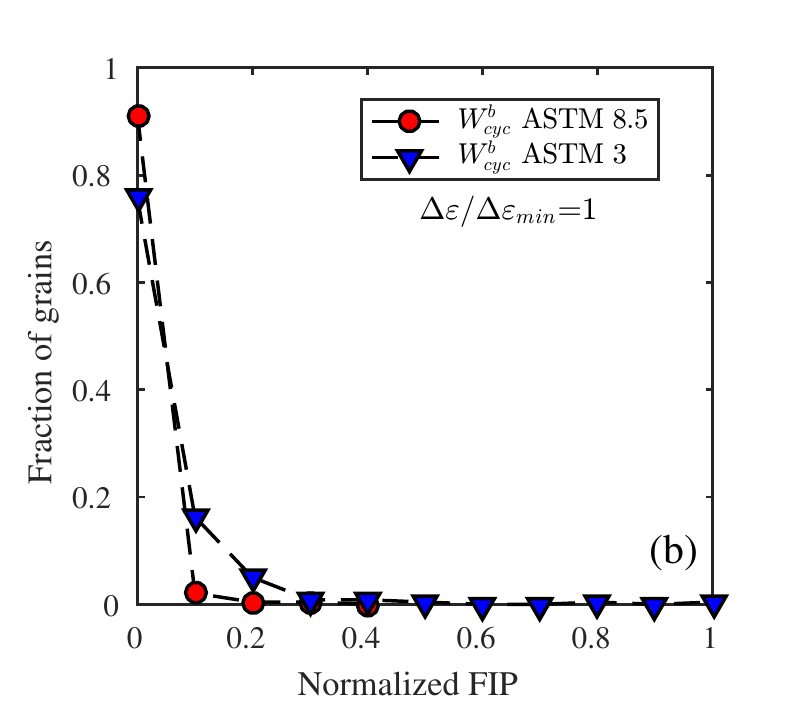}
\caption{Histogram of the stabilized, cyclic FIP in one RVE for the IN718 alloys with different grain sizes, ASTM 8.5 and ASTM 3. (a) $\Delta \varepsilon/\Delta \varepsilon_{min}$ = 3.5, (b)$\Delta \varepsilon/\Delta \varepsilon_{min}$ = 1. The FIP in each figure is normalized by the maximum FIP value in both microstructures.} 
\label{distri_fip}
\end{figure}

Moreover, it was found that the influence of the microstructure is very large in the case of localized deformation, because the fatigue life is controlled by a continuous accumulation of plastic slip in a few grains. The histograms of the distribution of the stabilized, cyclic FIP within the RVE are plotted in Fig. \ref{distri_fip}a)  and b) for $\Delta \varepsilon/\Delta \varepsilon_{min}$ =  3.5 and 1, respectively. The RVE selected to obtain the histogram corresponds to the one that gives the best prediction of the experimental results for both microstructures.  In each figure, the stabilized FIP value is normalized by the maximum FIP value in both microstructures. Plastic deformation was homogeneously distributed throughout the microstructure of both materials in the simulations carried out with $\Delta \varepsilon/\Delta \varepsilon_{min}$ = 3, leading to very similar FIP histograms (Fig. \ref{distri_fip}a) and to equivalent fatigue lives, regarding of the grain size. However, the FIP distribution is Fig. \ref{distri_fip}b)
shows that plastic deformation (and, thus, non-zero values of the FIP) was localized in a few grains of the microstructure for $\Delta \varepsilon/\Delta \varepsilon_{min}$ = 1.  Under these conditions, the effect of grain size is clearly visible: the maximum FIP in the alloy with large grain size (ASTM 3) was significantly higher than that in the alloy with small grain size (ASTM 8.5). Moreover, the fraction of grains in which the FIP $>$ 0 was higher in the alloy with large grain size. Both factors show the beneficial effect of the small grain size from the fatigue crack initiation viewpoint and explain the longest fatigue life of the alloy with small grain size in this regime.

\begin{figure}[H]
\centering
\includegraphics[scale=0.9]{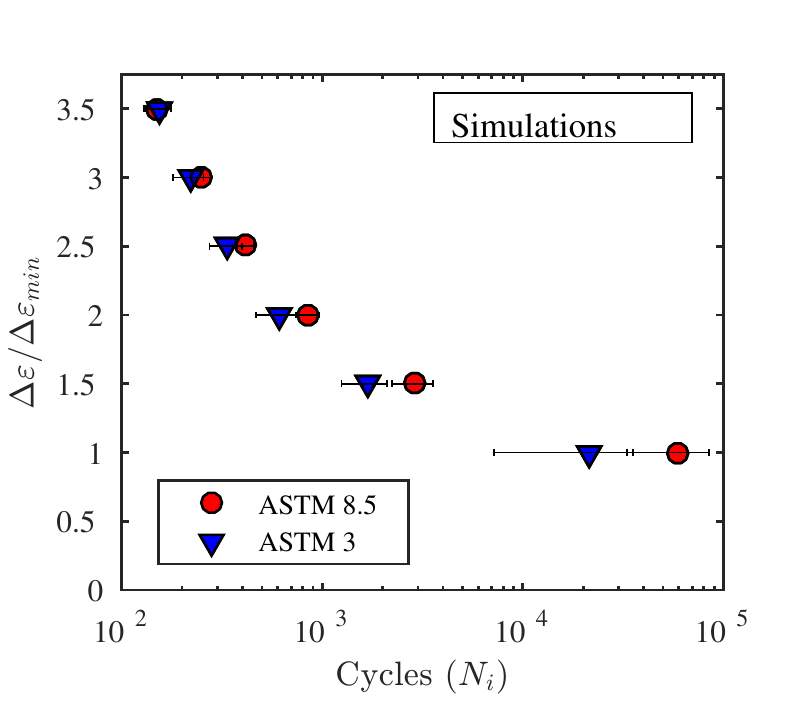}
\caption{Numerical predictions of the effect of the grain size in number of cycles for fatigue crack initiation (average and standard deviation) in IN718 alloys with grain sizes ASTM 8.5 and ASTM 3  at 400$^{\circ}$C as a function of the applied cyclic strain range, $\Delta \varepsilon$ (normalized by the minimum cyclic strain range, $\Delta \varepsilon_{min}$).} 
\label{fatigue_FEM_grain}
\end{figure}

In summary, the small differences in the plastic behavior induced by differences in the average grain size have no effect in the number of cycles for crack nucleation when plasticity is homogeneously distributed throughout the microstructure. However,  these small differences give rise to large differences in the fatigue life when plasticity is highly localized in a few grains. This leads to a large experimental scatter and is also responsible for the influence of the grain size on the fatigue life in this regime. This result is shown in Fig. \ref{fatigue_FEM_grain} where the average number of cycles for fatigue crack initiation (and the corresponding standard deviation) predicted for each cyclic strain range, $\Delta \varepsilon/\Delta \varepsilon_{min}$, is plotted for both microstructures. Accumulation of plastic deformation is homogeneous throughout the microstructure at large cyclic strain ranges and the scatter in the fatigue crack initiation predictions is low and the fatigue lives were independent of the grain size.  On the contrary,  plastic deformation is highly localized in a few grains and bands at small cyclic strain ranges. As a result, the scatter in the fatigue life  depends on the microstructural details and is much larger and the alloy with smaller grain sizes performs better in fatigue than the coarse grain counterpart in this regime.

Therefore, the fatigue life models \cite{Alexandre2004,Shenoy2007,Castelluccio2015} that take into account the effect of grain size effect following the mechanism proposed by Tanaka and Mura \citep{Tanakka1981} are not appropriate to explain the behavior of wrought IN718 alloys. These models assume that the plastic behavior of the grains is independent of grain size and that crack initiation is inversely proportional to the square of the plastic strain amplitude and of the grain size. Under these hypotheses, the ratio of the fatigue life of both microstructures (fine and coarse) should be constant regardless of the applied cyclic strain range  but this is not the case  in the experimental results of IN718 (Fig.  \ref{fatigue}). The origin of the breakdown of this proportionality in the fatigue life in IN718 is the redistribution of plastic microfields from homogeneously distributed to highly localized, which is also responsible of the dual slope in the Coffin-Manson plot for this alloy \cite{Cruzado2017b}.

\section{Conclusions} 

The fatigue behavior of two wrought IN718 alloy alloys with fine (ASTM 8.5) and coarse (ASTM 3) grain size was studied by means of unaxial cyclic deformation tests under fully-reversed deformation ($R_\varepsilon$ = -1) at 400$^\circ$C in the low cycle fatigue regime. It was found that the grain size did not influence the number of cycles for crack initiation for large applied cyclic strain ranges but the alloy with fine grain size presented much higher resistance to crack nucleation for small applied cyclic strain ranges. 

A microstructure-based model was proposed to simulate the effect of grain size on the number of cycles for crack nucleation in IN718 alloy at 400$ ^\circ$C subjected to low cycle fatigue. The model was based in the determination of the fatigue indicator parameter (based on the local crystallographic strain energy dissipated per cycle throughout the microstructure) by means of computational homogenization of an RVE of the microstructure. The mechanical response of the single crystals within the polycrystal was modelled by means of a phenomenological crystal plasticity model that was able to reproduce accurately the main cyclic characteristics of IN718 alloy, namely kinematic hardening, mean stress relaxation and cyclic softening. The model accounts for the effect of grain size on both the isotropic hardening/softening and on the kinematic hardening through the introduction of a dependency on the grain size (based on the Hall-Petch model)  on the critical resolved shear stress and the back stress. The crystal plasticity model parameters were calibrated from the experimental cyclic stress-strain curves at different cyclic strain ranges and the parameters of the fatigue crack initiation model were calibrated using the number of cycles for crack initiation obtained experimentally for the alloy with the fine grain size under two different strain ranges.

The results of the fatigue simulations were in good agreement with the experimental results in terms of the cyclic stress-strain curves and of the number of cycles for fatigue crack initiation. The model did not show any grain size effect on the fatigue life for large applied strain ranges but the predicted fatigue life was considerably longer in the case of alloy with fine grain size for small cyclic strain ranges. These differences were attributed to changes in the deformation modes between these regimes. Plastic deformation is homogeneously distributed throughout the microstructure at large cyclic  strain ranges and the influence of the details of the microstructure (including the average grain size) on the fatigue life is minimum. However, plasticity is localized in a few grains at low cyclic strain ranges and  the influence of the microstructure is very large in the case of localized deformation, because the fatigue life is controlled by a the continuous accumulation of plastic slip in a few grains. This leads to a large experimental scatter and is also responsible for the influence of the grain size on the fatigue life in this regime. 

\section*{Acknowledgments}

This investigation was supported by the European Union through the Clean Sky Joint Undertaking, 7th Framework Programme, project MICROMECH  (CS-GA-2013-620078) and by the Spanish Ministry of Economy and Competitiveness (project DPI2015-67667) and by the European Research Council under the European Union Horizon 2020 research and innovation programme (Advanced Grant VIRMETAL, grant agreement No. 669141). In addition, the authors thank Industria de TurboPropulsores. S. A. and, in particular,  Dr. A. Linaza and Dr. K. Ostolaza for providing the experimental data and their support during the project. The experimental data included in this paper are proprietary of the industrial partners in the MICROMECH project and, as a result, the stresses and strains have been normalized by a constant factor.


\end{document}